\documentclass[aps,prb,twocolumn,showpacs,preprintnumbers,amsmath,amssymb,superscriptaddress]{revtex4}%

\usepackage{graphicx}%
\usepackage{dcolumn}
\usepackage{amsmath}
\usepackage{color}

\begin{document}

\title{ Proximity induced superconductivity within the insulating (Li$_{0.84}$Fe$_{0.16}$)OH layers in (Li$_{0.84}$Fe$_{0.16}$)OHFe$_{0.98}$Se}
\author{Rustem~Khasanov}
 \email{rustem.khasanov@psi.ch}
 \affiliation{Laboratory for Muon Spin Spectroscopy, Paul Scherrer Institut, CH-5232 Villigen PSI, Switzerland}
\author{Huaxue~Zhou}
 \affiliation{Beijing National Laboratory for Condensed Matter Physics, Institute of Physics {\rm \&} University of Chinese Academy of Science, CAS, Beijing 100190, China}
\author{Alex~Amato}
 \affiliation{Laboratory for Muon Spin Spectroscopy, Paul Scherrer Institut, CH-5232 Villigen PSI, Switzerland}
\author{Zurab~Guguchia}
 \affiliation{Laboratory for Muon Spin Spectroscopy, Paul Scherrer Institut, CH-5232 Villigen PSI, Switzerland}
\author{Elvezio~Morenzoni}
 \affiliation{Laboratory for Muon Spin Spectroscopy, Paul Scherrer Institut, CH-5232 Villigen PSI, Switzerland}
\author{Xiaoli~Dong}
 \affiliation{Beijing National Laboratory for Condensed Matter Physics, Institute of Physics {\rm \&} University of Chinese Academy of Science, CAS, Beijing 100190, China}
\author{Guangming~Zhang}
 \affiliation{State Key Laboratory of Low-Dimensional Quantum Physics {\rm \&} Department of Physics, Tsinghua University, Beijing 100084, China}
\author{Zhongxian~Zhao}
 \affiliation{Beijing National Laboratory for Condensed Matter Physics, Institute of Physics {\rm \&} University of Chinese Academy of Science, CAS, Beijing 100190, China}

\begin{abstract}
The role played by the insulating intermediate (Li$_{0.84}$Fe$_{0.16}$)OH layer on magnetic and superconducting properties of (Li$_{0.84}$Fe$_{0.16}$)OHFe$_{0.98}$Se was studied by means of muon-spin rotation.  It was found that it is not only enhances the coupling between the FeSe layers for temperatures below $\simeq 10$~K, but becomes superconducting by itself due to the proximity to the FeSe ones. Superconductivity in (Li$_{0.84}$Fe$_{0.16}$)OH layers is most probably filamentary-like and the energy gap value, depending on the order parameter symmetry, does not exceed 1-1.5~meV.
\end{abstract}


\maketitle

Recently, the iron chalcogenide system has attracted much interest due to a series of discoveries of  new superconductors with high transition temperatures ($T_c$'s). $T_c$ of FeSe$_{1-x}$ reaches values up to $\simeq$ 37~K by applying pressure.\cite{Margadonna_PRB_09}
An intercalation of the alkali metals (K, Cs, Rb) between FeSe layers increases $T_c$ above 30~K. \cite{Guo_PRB_10, Krzton-Maziopa_JPCM_11, Wang_PRB_11} The $T_c$ value raises up to $\simeq$ 100~K in a single layer FeSe film grown on a SrTiO$_3$ substrate. \cite{Wang_CPL_12, Ge_NatM_15}
Significant enhancement of $T_c$ is also observed in FeSe structures intercalated with alkali metal coordinated to molecular spacers (as {\it e.g.}, ammonia, pyridine, ethylenediamine, or hexamethylenediamine) \cite{Ying_ScRep_12, Burrard-Lucas_NatM_12, Krzton-Maziopa_JPCM_12, Hosono_JPSJ_14} as well as by lithium-iron hydroxide. \cite{Lu_NatM_14, Pachmayr_ACIE_15, Sun_IC_15, Dong_JACS_15, Dong_PRB_15} In this case $T_c$ was also claimed to increase with the  increased two-dimensionality. \cite{Zhang_SciRep_13} However, to date, the enhancement of two-dimensional properties caused by intercalation were solely related to the increased distance between the superconducting FeSe layers. \cite{Hosono_JPSJ_14, Zhang_SciRep_13} Due to the lack of good quality single crystals, the anisotropic physical properties as well as the role of the intermediate spacer layer were not yet studied. Recently Dong {\it et al.} \cite{Dong_PRB_15} have reported the synthesis of high-quality single crystals of (Li$_{1-x}$Fe$_{x}$)OHFeSe with $T_c$ reaching $\simeq 42$~K. The highly anisotropic properties of (Li$_{1-x}$Fe$_{x}$)OHFeSe were confirmed in the measurements of the normal state resistivity and upper critical field.
In this letter we report on a detailed study of the evolution of the superconducting and magnetic properties of single crystalline (Li$_{0.84}$Fe$_{0.16}$)OHFe$_{0.98}$Se by using the muon-spin rotation ($\mu$SR) technique.%

The $\mu$SR experiments in zero-field (ZF-$\mu$SR) were performed in order to study the magnetic response of the sample. In two sets of experiments the initial muon-spin polarization $P(0)$ was applied parallel to the crystallographic $c-$axis and the $ab$ plane, respectively. 
Experiments confirm the homogeneity of the sample as well as the magnetic ordering within the (Li$_{0.84}$Fe$_{0.16}$)OH layer below $T_m\simeq10$~K (see the Supplementary information). Note that recent density functional theory calculations reveal that depending on the amount of Fe distribution disorder within the intermediate  (Li$_{1-x}$Fe$_{x}$)OH layer, both, the ferromagnetic and the antiferromagnetic type of order become possible. \cite{Chen_Arxiv_15} For the particular sample studied here, the absence of ``ferromagnetic-like" features on magnetization curves (see the Supplementary information and Ref.~\onlinecite{Pachmayr_ACIE_15}), the weak influence of magnetism on muons stopped within FeSe layers (see the Supplementary information) and the results of "field-shift" experiments (see below) suggest that the magnetic order in the intermediate (Li$_{0.84}$Fe$_{0.16}$)OH layer is antiferromagnetic like.

\begin{figure}[htb]
\includegraphics[width=0.9\linewidth]{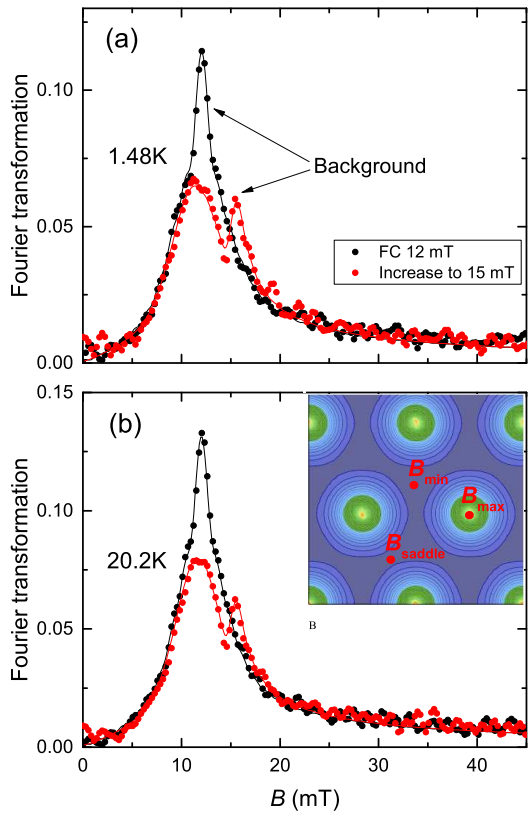}
%
\caption{Fast Fourier transform of TF-$\mu$SR time spectra after cooling in an applied field of $\mu_0H=12$~mT, $H\parallel c$, (black symbols) and after a subsequent field increase to 15~mT (red symbols) at $T=1.48$~K (a) and 20.2~K (b). The solid lines are fits (see the Supplementary Information). The inset in (b) is the contour plot of the field variation within the triangular vortex lattice. $B_{\rm min}$, $B_{\rm max}$, and $B_{\rm saddle}$ are the minimum, maximum and the saddle point fields.  }
 \label{fig:Field-Shift}
\end{figure}

The homogeneity of the superconducting state was checked by performing series of field-shift experiments in transverse-field (TF) configuration. Figure~\ref{fig:Field-Shift} exhibits the Fast Fourier transform  of the TF-$\mu$SR time spectra, which reflects the internal field distribution $P(B)$. The black curve corresponds to $P(B)$ obtained after cooling the sample at $\mu_0H=12$~mT ($H\parallel c-$axis) from a temperature above $T_c$ down to 1.48~K (Fig.~\ref{fig:Field-Shift}a) and 20.2~K (Fig.~\ref{fig:Field-Shift}b). The sharp peak at $B\simeq 12$~mT, accounting for approximately 8\% of the total signal amplitude is attributed to the residual background signal from muons missing the sample. The red curves are the $P(B)$ distributions after field cooling in 12~mT and subsequently increasing it up to 15~mT. The analysis reveals that the background signal shifts up by $\simeq3$~mT in agreement with the change of the applied field, whereas the signal from the sample remains unchanged within the experimental error. The asymmetric $P(B)$ distributions reported on Fig.~\ref{fig:Field-Shift} possess the basic features expected for a well aligned vortex lattice. In the case of triangular lattice (inset in Fig.~\ref{fig:Field-Shift}b) the cutoff at low fields corresponds to the minimum in $P(B)$ occurring at the midpoint of three adjacent vortices. The peak arises from the saddle point midway between two adjacent vortices, whereas the long tail towards high fields is due to the region around the vortex core.
The field-shift experiment clearly demonstrates that: (i) the vortex lattice in (Li$_{0.84}$Fe$_{0.16}$)OHFe$_{0.98}$Se sample is strongly pinned in a similar way above (20.2~K) and below (1.48~K) the magnetic ordering temperature ($T_m\simeq10$~K). Note that in the case of a ferromagnetically ordered (Li$_{1-x}$Fe$_{x}$)OH layer the pinning vanishes at $T\simeq1.5$~K;\cite{Pachmayr_ACIE_15}  (ii) the absence of any background peak in the unshifted signal implies that the sample is free of sizeable nonsuperconducting inclusions.

\begin{figure}[htb]
\includegraphics[width=0.9\linewidth]{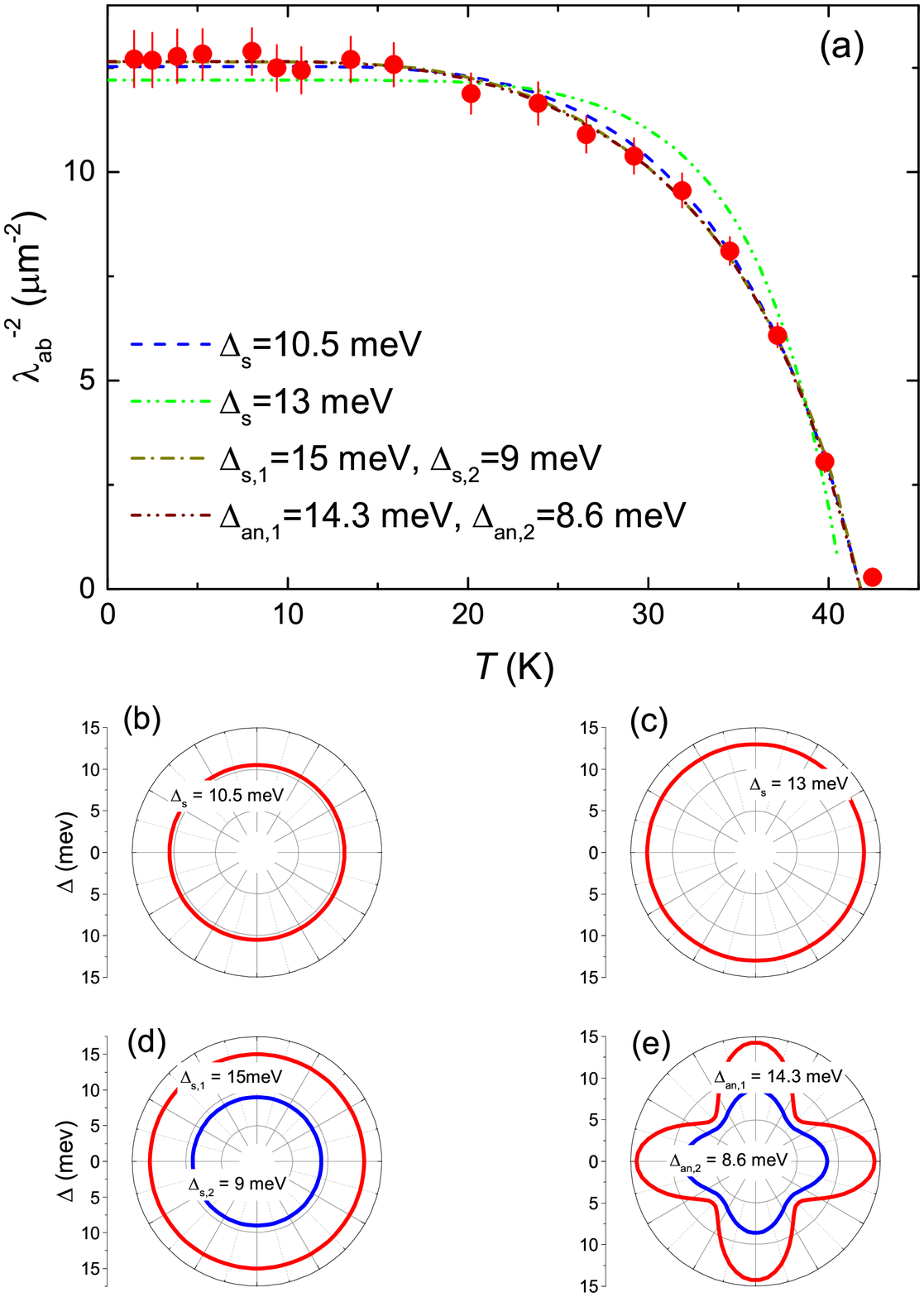}
%
\caption{ (a) Temperature dependence of $\lambda_{ab}^{-2}$ of
(Li$_{0.84}$Fe$_{0.16}$)OHFe$_{0.98}$Se. The fitting curves were obtained within the following
models of gap symmetries: $s-$wave: $\Delta_s=10.5$~meV (Ref.~\onlinecite{Niu_PRB_ARPES_15} and panel b); $s-$wave: $\Delta_s=13$~meV (Ref.~\onlinecite{Zhao_Arxiv_ARPES_15} and panel c);
two $s-$wave gaps: $\Delta_{s,1}= 15$~meV and $\Delta_{s,2}= 9$~meV (Ref.~\onlinecite{Yan_arxiv_STS_15} and panel d); and two anisotropic gaps: $\Delta_{an,1}=14.3$~meV$\cdot[1-0.25(1-\cos4\varphi)]$ $\Delta_{an,2}=8.6$~meV$\cdot[1-0.15(1-\cos4\varphi)]$ (Ref.~\onlinecite{Du_Arxiv_STS_15} and panel e).}
 \label{fig:Lambda_ab}
\end{figure}

\begin{figure}[htb]
\includegraphics[width=0.9\linewidth]{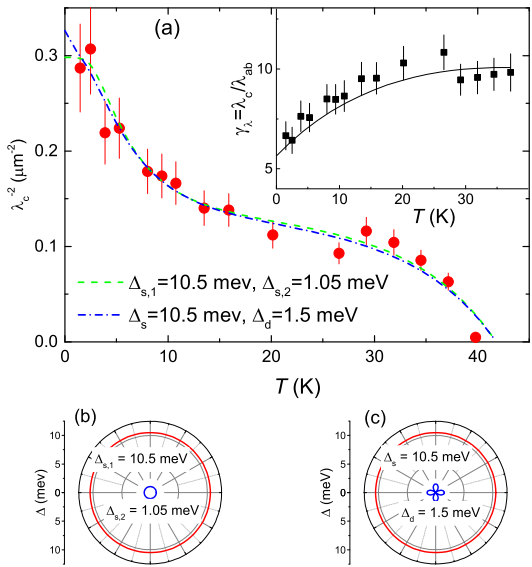}
%
\caption{ (a) Temperature dependence of $\lambda_{c}^{-2}$ of
(Li$_{0.84}$Fe$_{0.16}$)OHFe$_{0.98}$Se. The fitting curves were obtained within the $s+s$ (b) and $s+d$ (c) gap symmetries. The larger gap and $T_c$ were fixed to $\Delta_s=10.5$~meV and $T_c=41.9$~K, respectively. The inset in (a) shows the temperature dependence of the anisotropy parameter $\gamma_\lambda=\lambda_c/\lambda_{ab}$. The line is a guide for the eye. }
 \label{fig:Lambda_c}
\end{figure}

The temperature dependences of the in-plane ($\lambda_{ab}$) and the out-of-plane ($\lambda_c$) components of the magnetic penetration depth were studied in TF-$\mu$SR experiments with the external magnetic field applied parallel to the $c-$axis and the $ab-$plane, respectively. The details of the data analysis procedure are presented in the Supplementary Information. We just recall here that the inverse squared penetration depth is directly proportional to the superfluid density $\lambda^{-2}\propto\rho_s$. The corresponding dependences of $\lambda_{ab}^{-2}\propto\rho_{s,ab}$ and $\lambda_{c}^{-2}\propto\rho_{s,c}$ on $T$ are shown in Figs.~\ref{fig:Lambda_ab}~a and \ref{fig:Lambda_c}~a.

As a first step, we discuss the temperature dependence of $\lambda_{ab}^{-2}$ since it is primarily determined by the superconducting gap(s) opening in the $ab$ plane. They should correspond to the ones measured directly in recent angle-resolved photoemission (ARPES) and scanning tunneling spectroscopy (STS) experiments. \cite{Niu_PRB_ARPES_15, Zhao_Arxiv_ARPES_15, Du_Arxiv_STS_15, Yan_arxiv_STS_15}  At present, there is no consistency on the number of gaps (one versus two), their symmetries, and absolute values. ARPES experiments reveal the presence of a single band around the $M$ point at the Brillouin zone with an isotropic ($s-$wave) gap. The reported gap value varies between $\Delta_s=10.5$~meV and $\simeq 13$~meV.\cite{Niu_PRB_ARPES_15,Zhao_Arxiv_ARPES_15} STS experiments points to presence of two sets of electron pockets near the $M$ point with different symmetries and high values of the gaps. Two $s-$wave gaps with $\Delta_{s,1}\simeq 15$~meV and $\Delta_{s,2}\simeq 9$~meV were found in Ref.~\onlinecite{Yan_arxiv_STS_15}, while two anisotropic gaps with maximum values $\Delta_{an,1}=14.3$~meV and $\Delta_{an,2}=8.6$~meV were observed in Ref.~\onlinecite{Du_Arxiv_STS_15}. The angular distributions of gaps reported in Refs.~\onlinecite{Niu_PRB_ARPES_15, Zhao_Arxiv_ARPES_15, Du_Arxiv_STS_15, Yan_arxiv_STS_15} are shown schematically in Figs.~\ref{fig:Lambda_ab} (b)--(e).

The temperature dependence of $\lambda_{ab}^{-2}$ was further analyzed within the local (London) approach
by using the following functional form: \cite{Tinkham_75, Khasanov_PRL_La214_07}
\begin{equation}
\frac{\lambda^{-2}(T)}{\lambda^{-2}(0)}=  1
+\frac{1}{\pi}\int_{0}^{2\pi}\int_{\Delta(T,\varphi)}^{\infty}\left(\frac{\partial
f}{\partial E}\right)\frac{E\
dEd\varphi}{\sqrt{E^2-\Delta(T,\varphi)^2}}~.
 \label{eq:lambda-d}
\end{equation}
Here $\lambda^{-2}(0)$ is the zero-temperature value  of the magnetic penetration depth, $f=[1+\exp(E/k_BT)]^{-1}$ is  the Fermi function, $\varphi$ is the angle along the Fermi surface,
and $\Delta(T,\varphi)=\Delta \ g(\varphi) \tanh\{1.82[1.018(T_c/T-1)]^{0.51}\} $ [$\Delta$ is the gap value at $T=0$].\cite{Khasanov_PRL_La214_07}  $g(\varphi)$ describes the angular dependence of the gap: $g_s(\varphi)=1$ for the $s-$wave gap, $g{_{an}}(\varphi)=[1-a(1-\cos 4\varphi)]$ for the anisotropic gap, \cite{Du_Arxiv_STS_15} and $g_d(\varphi)=|\cos(2\varphi)|$ for the $d-$wave gap.

The two-gap analysis was performed within the  framework of the phenomenological $\alpha-$model: \cite{Khasanov_PRL_La214_07, Carrington_03}
\begin{equation}
\frac{\lambda^{-2}(T)}{\lambda^{-2}(0)}=
\omega\; \frac{\lambda^{-2}(T,
\Delta_{1})}{\lambda^{-2}(0,\Delta_{1})}+(1-\omega)\;
\frac{\lambda^{-2}(T, \Delta_{2})}{\lambda^{-2}(0,\Delta_{2})}.
 \label{eq:two-gap}
\end{equation}
Here $\omega$ ($0\leq\omega\leq1$) is the weight factor representing the relative contribution of the larger gap to  $\lambda^{-2}$.

The results of the analysis are presented  in Fig.~\ref{fig:Lambda_ab}a and Table~I in the Supplementary Information. It should be noted here that the fits were performed by using the gap values and the gap symmetries as measured in ARPES and STS experiments (see Refs.~\onlinecite{Niu_PRB_ARPES_15, Zhao_Arxiv_ARPES_15, Du_Arxiv_STS_15, Yan_arxiv_STS_15} and also Figs.~\ref{fig:Lambda_ab} b--e).  The only free parameters were $\lambda_{ab}^{2}(0)$ and $T_c$ in the case of single $s-$wave gap fits and  $\lambda_{ab}^{2}(0)$, $T_c$, and $\omega$ within a two-gap approach.
Obviously, three out of four gap models describe the obtained $\lambda_{ab}^{-2}(T)$ dependence almost equally well. Only the curve with $\Delta_s=13$~meV deviates significantly from the data.

Two important points need to be considered: (i) The analysis reveals that within the single $s-$wave gap approach a satisfactory agreement between the fit and the data is achieved for $9.8\lesssim \Delta_s\lesssim 10.6$~meV. The gap value of 10.5~meV measured in Ref.~\onlinecite{Niu_PRB_ARPES_15} stays within this limit, while the value of 13~meV  from Ref.~\onlinecite{Zhao_Arxiv_ARPES_15} is $\simeq 15$\% higher. (ii) As shown in  Table~I in the Supplementary Information, in the case of two anisotropic gaps, the relative weight of the smaller gap is consistent with zero. This suggests that the bands where the smaller gap is supposed to open do not supply {\it any} supercarriers to the superfluid density and, consequently, the energy gap cannot exist. However, the analysis reveals that by decreasing the degree of the larger gap anisotropy ($a=0.25$, Ref.~\onlinecite{Du_Arxiv_STS_15}), the weight of the smaller gap continuously increases, reaching $\simeq 40$\% for $a=0$. This implies that if the two anisotropic gaps scenario is realized in (Li$_{1-x}$Fe$_{x}$)OHFeSe, the larger gap should have a smaller anisotropy than suggested in Ref.~\onlinecite{Du_Arxiv_STS_15}.

The temperature dependence of the inverse squared out-of-plane magnetic penetration depth $\lambda^{-2}_c\propto\rho_{s,c}$ is shown in Fig.~\ref{fig:Lambda_c}. It is reasonable to assume that the superconducting energy gap(s) detected within the $ab$ plane should remain the same in the perpendicular direction. This seems to be correct for some Fe-based superconductors as {\it e.g.} SrFe$_{1.75}$Co$_{0.25}$As$_2$, \cite{Khasanov_PRL_09} FeSe$_{0.5}$Te$_{0.5}$, \cite{Bendele_PRB_10}  LiFeAs, \cite{Song_EPL_11} which are characterized by relatively small values of the anisotropy parameter  $\gamma_{\lambda}=\lambda_c/\lambda_{ab}$. By lowering the temperature $\gamma_{\lambda}$ changes from $\gamma_\lambda\simeq 2.0$, 1.5 and 2.0 close to $T_c$ to $\gamma_\lambda\simeq 2.7$, $2.5$ and 1.0 at $T\simeq0$  for SrFe$_{1.75}$Co$_{0.25}$As$_2$,  FeSe$_{0.5}$Te$_{0.5}$ and LiFeAs, respectively. \cite{Khasanov_PRL_09, Bendele_PRB_10,Song_EPL_11}

The analysis reveals, however, that {\it none} of the gap models describing $\lambda_{ab}^{-2}(T)$ agree with the $\lambda_c^{-2}(T)$ dependence. The inflection point at $T\simeq 10$~K clearly implies that a superconducting gap with an absolute value much smaller than determined by means of ARPES and STS is present.  Bearing this in mind and accounting for the simplest $s-$wave model describing $\lambda_{ab}^{-2}(T)$ ($\Delta_s=10.5$~meV, $T_c=41.9$~K, see Fig.~\ref{fig:Lambda_ab} and Table~I in the Supplementary Information), the two-gap model (Eq.~\ref{eq:two-gap}) with the larger gap $\Delta_s=10.5$~meV and the smaller gap remaining as a free parameter was fitted to the $\lambda_c^{-2}(T)$.
For the smaller gap the $s-$wave and $d-$wave type of symmetries were considered. The results of the fit are presented in Fig.~\ref{fig:Lambda_c}~a and Table~~I in the Supplementary Information. The angular distributions of the gaps in the case of $s+s$ and $s+d$ model fittings are shown schematically in Figs.~\ref{fig:Lambda_c}~b and c.
Both, $s+s$ and $s+d$, gap models fit $\lambda_c^{-2}(T)$ equally well. One cannot distinguish between them within the accuracy of the experiment. The gap values ($\Delta_{s,2}=1.05$~meV and $\Delta_d=1.50$~meV) are factor of 5 to 10 lower than the smallest gap within the $ab$ plane. This clearly differentiates (Li$_{1-x}$Fe$_{x}$)OHFeSe from other Fe-based superconductors where the gap(s) were found to be essentially direction independent.

In (Li$_{1-x}$Fe$_{x}$)OHFeSe the FeSe layers are weakly bonded to the intermediate (Li$_{1-x}$Fe$_x$)OH layers via hydrogen atoms. The distance between the superconducting  FeSe layers ($\simeq 9.3$~\AA) is also higher than in most of the Fe-based superconducting families (including, {\it e.g.} 11, 111, 122, and 245 families). This clearly indicates that (Li$_{1-x}$Fe$_{x}$)OHFeSe has a highly two-dimensional character, which is also confirmed experimentally by:  (i) the observation of an extremely high resistivity ratio $\rho_c/\rho_{ab}$ which increases continuously with decreasing temperature by reaching the value of $\rho_c/\rho_{ab}\simeq 2500$ at $T=50$~K;\cite{Dong_PRB_15} (ii) The similar electronic structure of (Li$_{1-x}$Fe$_{x}$)OHFeSe and the single-layer FeSe on the SrTiO$_3$ substrate observed by means of ARPES and STS. \cite{Niu_PRB_ARPES_15, Zhao_Arxiv_ARPES_15, Du_Arxiv_STS_15, Yan_arxiv_STS_15} Both systems were found to have a similar Fermi-surface topology, band structure and superconducting gap symmetry. This statement is further confirmed by an agreement of our $\lambda_{ab}^{-2}(T)$ data with the recent gap measurements from Ref.~\onlinecite{Zhang_arxiv_ARPES_15}, see the Supplementary Information; (iii) The present observation of a high value of the magnetic field penetration depth anisotropy. $\gamma_\lambda=\lambda_c/\lambda_{ab} \simeq 10$ close to $T_c$ which decreases to $\simeq 7$ at $T=1.5$~K (see the inset in Fig.~\ref{fig:Lambda_c}~a).

We believe therefore, that the enhanced two-dimensionality of (Li$_{1-x}$Fe$_{x}$)OHFeSe leads to the unusual observation of smaller gap opening along the crystallographic $c-$direction.  The large $s-$wave gap(s) (or large anisotropic gaps, see Table~~I in the Supplementary Information) correspond to a condensation of the supercarriers confined within the two-dimensional FeSe layers, whereas the tiny small gap opens in the (Li$_{1-x}$Fe$_x$)OH layers and is induced by the superconducting FeSe layers due to proximity effects.
Such situation is similar to the appearance of a proximity-induced gap in CuO chains  in the cuprate superconductor YBa$_2$Cu$_4$O$_8$. \cite{Kondo_PRL_10}   The superconducting gap detected in the chains ($\simeq 5$~meV) is significantly smaller than the gap in the superconducting CuO$_2$ planes ($\simeq 20$~meV) \cite{Khasanov_JSNM_08} and is confined within a very narrow $k-$space region. We should stress, however, on the significant difference between (Li$_{1-x}$Fe$_{x}$)OHFeSe studied here and YBa$_2$Cu$_4$O$_8$. In the later compound the chains are metallic and, therefore, allow for conductivity (superconductivity) along the chain direction. In (Li$_{1-x}$Fe$_{x}$)OHFeSe the resistivity anisotropy increases with decreasing temperature  \cite{Dong_PRB_15} thus suggesting that the intermediate (Li$_{1-x}$Fe$_x$)OH layers become more insulating.  Consequently, instead of expecting that the entire (Li$_{1-x}$Fe$_x$)OH remains conducting (superconducting), there is a lack of conducting channels between the FeSe layers through the (Li$_{1-x}$Fe$_x$)OH ones.

Indeed, there are indications from recent ARPES and STS experiments supporting the validity of the above scenario. STS measurements reveal that the (Li$_{1-x}$Fe$_x$)OH surface has a metallic behavior when the FeSe layers exhibit superconductivity. \cite{Yan_arxiv_STS_15} The tunnelling spectrum has a weak dip at a Fermi level which may point to a superconducting gap opening. We note that the modulation amplitude in STS experiments, $\Delta V=1$~meV,  is comparable to the values of the smallest gap obtained in our studies.  ARPES experiments on a similar (Li$_{1-x}$Fe$_{x}$)OHFeSe sample show that in addition to the electron-like bands crossing the Fermi level at around the $M$ point, there is a tiny electron-like weight at the Fermi energy near the $\Gamma$ point, which could be a contribution from the (Li$_{1-x}$Fe$_x$)OH layers.\cite{Zhao_Arxiv_ARPES_15}  The presence of such a tiny electron spectral weight might play a crucial role for the proximity effect.

We want to emphasize that the enhancement of the out-of-plane superfluid density ($\rho_{s,c}\propto\lambda_{c}^{-2}$) occurs at the same temperature range where the antiferromagnetic  order within the (Li$_{1-x}$Fe$_{x}$)OHFeSe layers sets in. The theory calculations presented in the Supplementary Information reveal that  carriers within the superconducting FeSe layers are strongly hybridized with the local Fe moments in (Li$_{1-x}$Fe$_x$)OH. The effect of such hybridization is two fold. First of all, it enhances the superconductivity within the FeSe layers, and, secondly, weak superconductivity with dominant $d-$wave symmetry can be induced in the insulating (Li$_{1-x}$Fe$_x$)OH layers. Note that the latter statement is consistent with $\Delta_d\simeq 1.5$~meV gap obtained from the fit of $\lambda_c^{-2}(T)$ data (see Fig.~\ref{fig:Lambda_c}).

In conclusion, the magnetic and superconducting properties of (Li$_{0.84}$Fe$_{0.16}$)OHFe$_{0.98}$Se single crystal were studied by means of muon-spin rotation technique. The zero-field and field-shift $\mu$SR experiments confirm the homogeneity of the sample and the antiferromagnetic ordering within the (Li$_{0.84}$Fe$_{0.16}$)OH layers below $T_m\simeq10$~K. The temperature dependences of the in-plane ($\lambda_{ab}$) and the out-of-plane ($\lambda_{c}$) components of the magnetic penetration depth were measured in transverse field $\mu$SR experiments. $\lambda_{ab}^{-2}(T)$ was found to be consistent with gap opening within the superconducting FeSe planes and it is well described within either the single $s-$wave gap ($\Delta_s=10.5$~meV) or two $s-$wave gaps ($\Delta_{s,1}=15$~meV and $\Delta_{s,2}=9$~meV) scenario in agreement with the results of ARPES \cite{Niu_PRB_ARPES_15} and STS \cite{Yan_arxiv_STS_15} experiments, respectively. The opening of an additional small superconducting gap of unknown symmetry ($\simeq$1.05~meV and 1.5~meV in a case of $s-$ and $d-$wave symmetry, respectively) was detected from $\lambda_{c}^{-2}(T)$.  This gap, most probably, opens within the insulating (Li$_{1-x}$Fe$_x$)OH layers and appears to be induced by the proximity to  the superconducting FeSe layers.
The strong enhancement of the out-of-plane superfluid density $\rho_{s,c}\propto\lambda_{c}^{-2}$ occurs at the same temperatures where the magnetism sets in.
The question on whether or not the superconductivity and antireferromagnetism within the intermediate (Li$_{1-x}$Fe$_x$)OH layers relate to the each other requires further studies.

This work was performed at the S$\mu$S Paul Scherrer Institute (PSI, Switzerland). The work of HZ, XD, and ZZ was supported by "Strategic Priority Research Program (B)" of the Chinese Academy of Sciences (No. XDB07020100) and Natural Science Foundation of China (projects 11574370 and 11274358 ). The work of ZG was supported by the Swiss National Science Foundation.

\end{document}